# Controlled spin-torque driven domain wall motion using staggered magnetic wires

H. Mohammed,[1] S. Al Risi,[2] T. L. Jin,[3] J. Kosel,[1] S. N. Piramanayagam,[3,a)] and R. Sbiaa[2,b)]

## AFFILIATIONS

[1]King Abdullah University of Science and Technology, Computer Electrical and Mathematical Science and Engineering, Thuwal 23955-6900, Saudi Arabia
[2]Department of Physics, Sultan Qaboos University, P.O. Box 36, PC 123 Muscat, Oman
[3]School of Physical and Mathematical Sciences, Nanyang Technological University, 21 Nanyang Link, 637371 Singapore

[a)]prem@ntu.edu.sg
[b)]rachid@squ.edu.om

## ABSTRACT

Domain wall (DW) memory devices such as racetrack memory offer an alternative to the hard disk drive in achieving high capacity storage. In DW memory, the control of domain wall positions and their motion using spin-transfer torque is an important challenge. In this Letter, we demonstrate controlled domain wall motion using spin-transfer torque in staggered wires. The devices, fabricated using electron-beam and laser lithography, were tested using magneto-optical Kerr microscopy and electrical transport measurements. The depinning current is found to depend on the device dimensions of the staggering wires. Thus, the proposed staggering configuration can be utilized to fine-tune the properties of DW devices for memory applications.

Spin-transfer torque (STT) has been intensively investigated as a tool to reverse the magnetization in a nanoscopic magnet,[1–12] generate high-frequency waves in nano-oscillators,[13–20] and move domain walls (DWs) in magnetic wires.[21–31] Although tremendous efforts have been dedicated to the reversal of magnetization in magnetic materials by the STT effect for magnetic random access memory (MRAM) applications, the storage capacity remains a serious challenge. The racetrack type domain wall memory (DWM) has the potential to combine the best performance of conventional MRAM in addition to the storage capacity.[32]

One of the major problems for adopting magnetic domain wall memory is the accuracy of moving domain walls (DWs) in precise positions. When recording each fresh bit of information onto a racetrack, there is considerable uncertainty about where each magnetic domain starts and ends, and an incorrectly written bit can easily lead to the corruption of all subsequent bits on the racetrack. To overcome these challenges, several ideas were proposed and tested, such as relying on physically pinning DWs.[33–41]

In a previous study, we demonstrated the possibility of stabilizing a domain wall in well-defined positions made in stepped wires.[36,37] However, the DW motion was investigated in materials with in-plane anisotropy using a magnetic field and without electric current. The displacement of a DW by STT effect is a better approach for the practical implementation of devices for commercial use.[25]

In this study, we demonstrate that a DW can be accurately moved by a polarized current in (Co/Ni) multilayers with perpendicular magnetic anisotropy. These magnetic multilayer structures have much higher magnetic anisotropy energy than in-plane types such as NiFe or CoFe alloy ferromagnets and, consequently, have better thermal stability.[42]

The magnetic stack consists of Ta(4)/Pt(2)/$[Co(0.3)/Ni(0.6)]_{\times 12}$/Pt(2)/Ta(2) deposited by DC-sputtering on a thermally oxidized Si substrate. The numbers show the thickness of each sub-layer in nanometers. The magnetic wire devices were fabricated using two lithography systems, namely, electron beam lithography (EBL) and direct-write laser (DWL) lithography. The EBL system was used to pattern wires along with nucleation pads on one end, and the DWL system was used to pattern Au electrodes onto the ends of the wires (Fig. 1). Figure 1(a) shows two small conventional wires with offsets in $x$ and $y$ directions to create a stepped device. The design of a multi-stepped wire with a 50 $\mu$m length and 1 $\mu$m width is shown in Fig. 1(b). The pad for nucleating magnetic domains is imaged with a scanning electron microscope (SEM). The whole device, including the electrical pad for injecting current, is also imaged by SEM in Fig. 1(d).

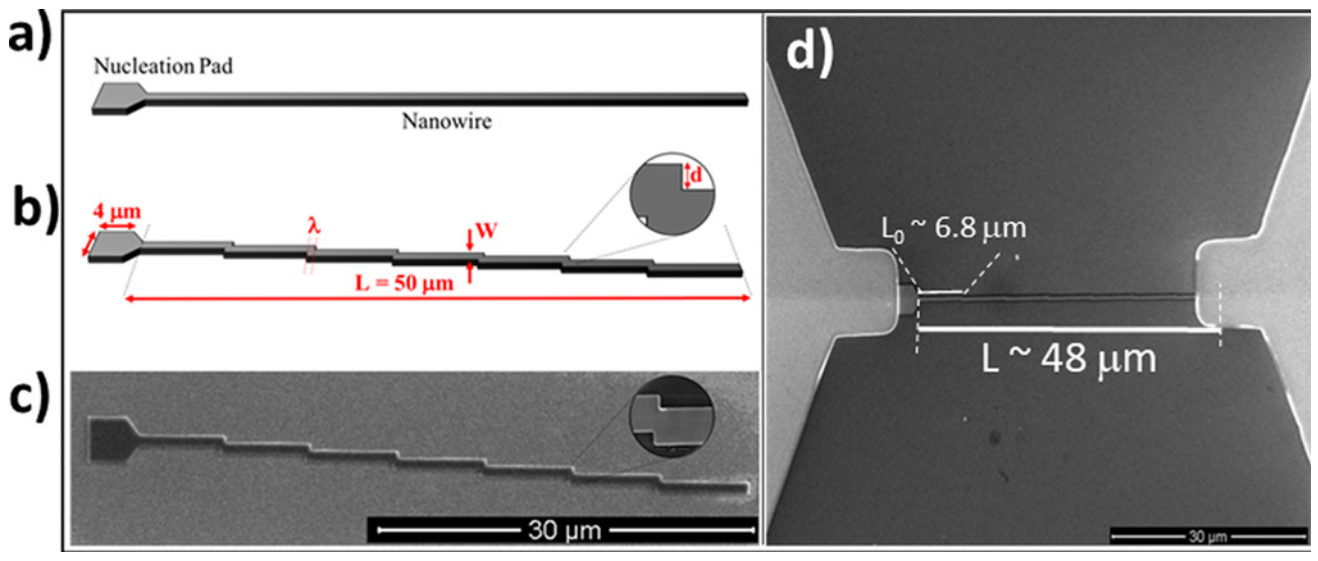

FIG. 1. The design of (a) a conventional and (b) a staggered magnetic wire with the $x$ and $y$ offsets represented by $\lambda$ and d, respectively. (c) Magnetic wire with multi-step design before making the two electrodes. The length and width of the wire were fixed to about 50 $\mu$m and 1 $\mu$m, respectively. (d) Scanning electron micrograph image of a part of the fabricated device after making the electrodes for applying the electric current.

The (Co/Ni) multilayers with 12 bilayers were first investigated as thin films using a physical property magnetometry system (PPMS) in out-of-plane geometry. In a prior study on (Co/Ni) multilayers, it was reported that there is a change of the shape of the hysteresis loop as the number of bilayers increases.[42] In this study, we selected the multilayer of 12 bilayers, which has a perpendicular magnetic anisotropy and small magnetic domains. The tail in the hysteresis loop shown in Fig. 2(a) is the result of an increase in magnetostatic energy, which competes with the magnetic anisotropy energy. The hysteresis loop for (Co/Ni) multilayers with eight bilayers is also plotted for comparison. The presence of the perpendicular magnetic anisotropy is confirmed by magnetic force microscopy (MFM) images shown in Figs. 2(b) and 2(c) for the cases of 8 and 12 bilayers, respectively. It is clear from MFM images that the size of the magnetic domains is strongly reduced for the sample with 12 bilayers as compared to that with 8. All the measurements were carried out at room temperature. Although the saturation of the unpatterned sample requires an external magnetic field of less than 1 kOe, after patterning the wires, the magnetization reversal occurs at a much higher field. To be able to observe the motion of a domain wall in wires, shown in Fig. 1(d), under a reasonable magnetic field (less than 1 kOe), the samples were annealed at 240 °C for 30 min. For the as-deposited samples, we could see a motion of the DW after applying a high external magnetic field of 0.9 T.

The procedure of investigating domain wall motion and magnetic reversal in a wire is schematically shown in Fig. 3(a). The device is under an external perpendicular magnetic field ($H_{per}$), and the pulsed current is applied along the wire. Simultaneously, a magneto-optical Kerr effect (MOKE) microscope is positioned above the device to image the magnetic domains. For better contrast, the recorded image was subtracted from a reference one, which is taken at the saturated state. The hysteresis loop shown in Fig. 3(b) is the result of magnetization reversal in a conventional wire without nano-constriction but with the same dimensions as the one shown in Fig. 1(b). For a conventional wire (without constriction), a swift change of the magnetic state from the up-state (state 1) to the down-state (state 2) was observed. Even with a very slow scan of the magnetic field, it was not possible to stabilize a DW at any position within the wire with the dimension discussed above. In the second step of this study, we investigated the pinning of the DW in stepped wires. The hysteresis loop of the wire is plotted in Fig. 4(a). Clear jumps in magnetization reversal could be seen, and the magnetic states were captured by MOKE microscopy. The first jump, which occurs at around 500 Oe, indicates the creation of a DW at the nucleation pad, stopped at the first step, as shown by the MOKE image in the inset. The second jump at a slightly higher applied magnetic field is the result of the creation of a magnetic domain from the right side of the wire. It has to be pointed out that this magnetic state is not due to the propagation of the first created domain on the left side. It is believed that the pinning strength at the constricted area is higher than the magnetic field needed for domain nucleation on the right side. It is important to note that the magnetic field from the electromagnet is covering the whole device area, and as a result, the nucleation may happen at any part of the device. Finally, at a larger magnetic field, the entire device magnetization becomes aligned in the up direction. There are a few steps where the DW could not be pinned or blocked.

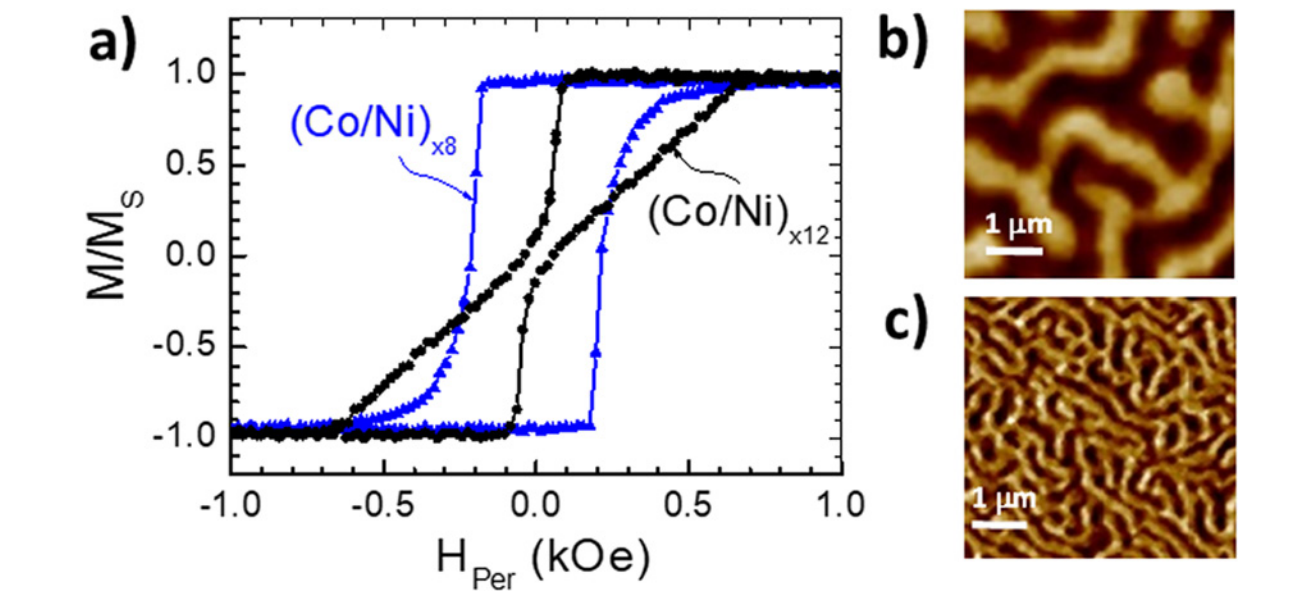

FIG. 2. (a) Hysteresis loops for thin film of a (Co/Ni) multilayer thin film with a perpendicular-to-plane applied magnetic field for 8 and 12 bilayers. Magnetic force microscopy image of (Co/Ni) multilayer with (b) 8 bilayers and (c) 12 bilayers at demagnetized state.

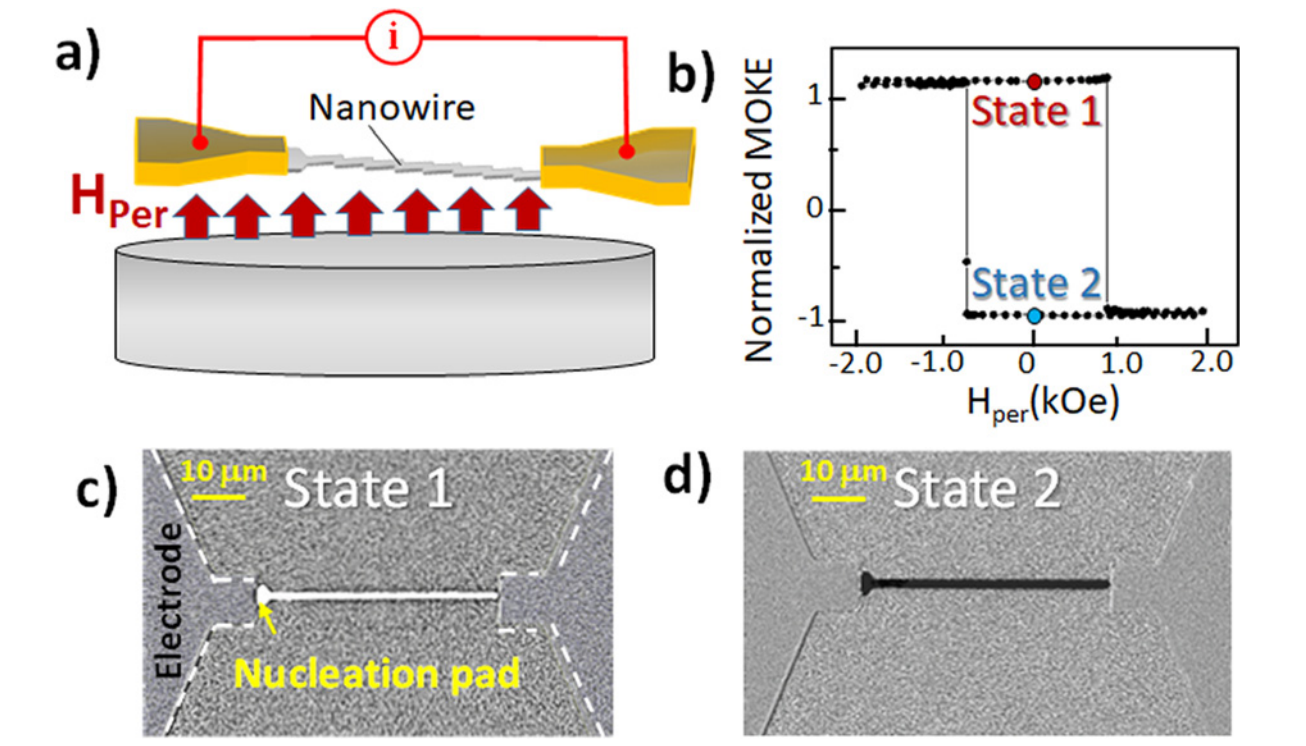

FIG. 3. (a) Schematic of the measurement setup, (b) polar MOKE signal for a conventional wire (without steps) showing a sharp change of MOKE signal at about 0.9 kOe applied magnetic field. An indication of only two possible states: (c) state 1 and (d) state 2 shown in (b).

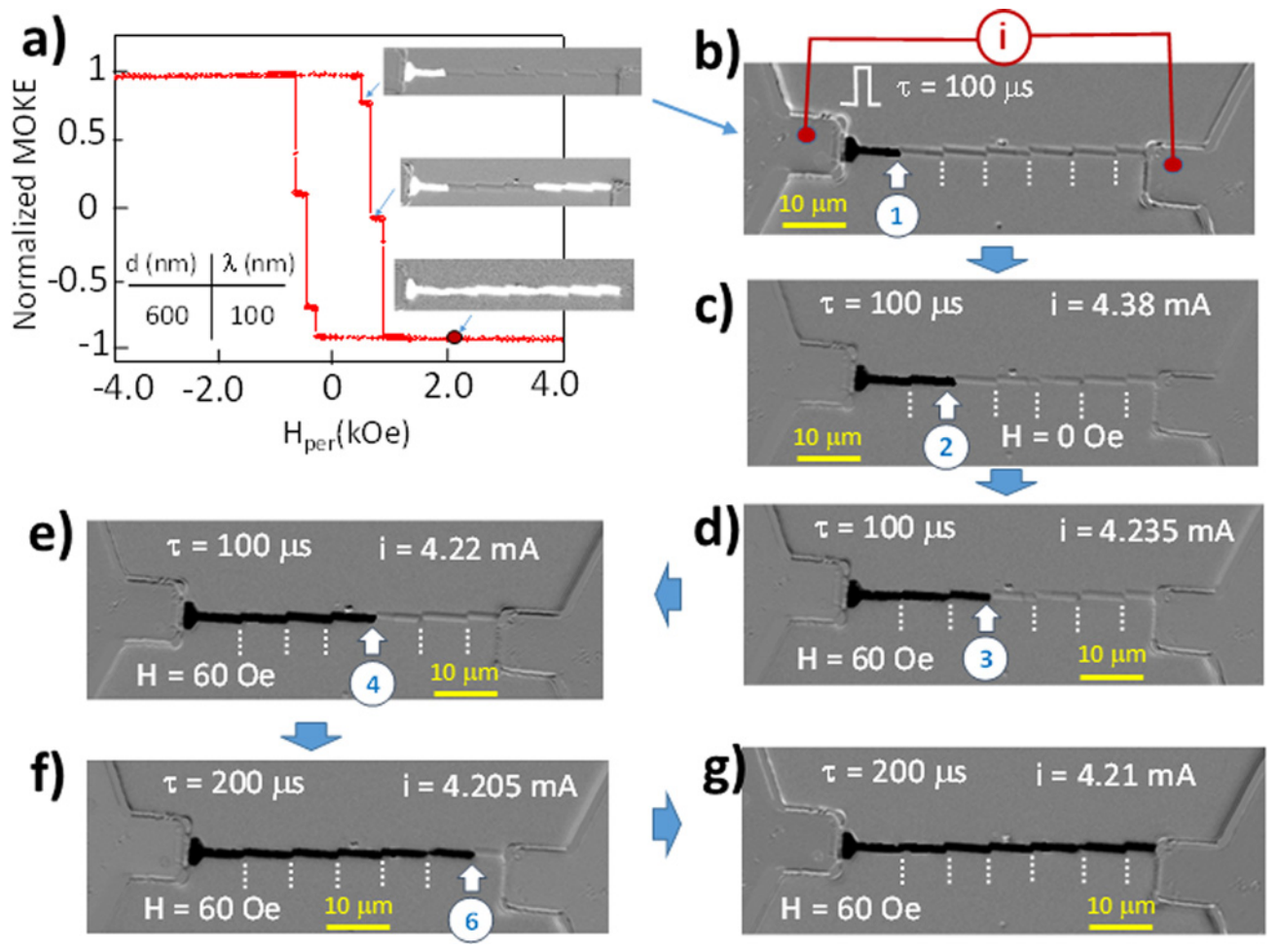

FIG. 4. The polar MOKE signal for a staggered wire as shown in Fig. 1(b). The insets are the MOKE images of different magnetic states at different applied magnetic fields. (b) MOKE image where the domain wall is stabilized by a magnetic field as the first step. (c)–(g) Other states obtained by applying different current magnitudes. The current is flowing along the wire, while the magnetic field is applied perpendicular to the device plane.

For DW motion by a magnetic field, good control of the strength of the effective pinning field in the stepped region is necessary. This can be achieved by intrinsic materials properties governing the domain wall dynamics such as anisotropy field, saturation magnetization, and exchange stiffness or by the device geometry or edge roughness. In this study, the fabricated devices have the same step dimensions, i.e., the same values of $d$ and $\lambda$ shown in Fig. 1(a). As a result, the pinning field offered by each step is the same. Therefore, despite our efforts, we were not successful in stabilizing a DW at each step by a magnetic field in several devices.

As our objective was to stabilize a DW at each step in the designed device, we carried out a similar study by applying an electric current between the two electrodes, as shown in Fig. 4(b) for values of $d$ and $\lambda$ of 600 nm and 100 nm, respectively. To start this series of experiments, we first created a magnetic domain near the nucleation pad by applying a small magnetic field in the out-of-plane direction, as shown in Fig. 3(a). The white arrow in Fig. 4(b) indicates the first step within the wire from the left side, where a DW is stabilized, and the small dotted lines show the other nano-constrictions for reference. To investigate the motion of the DW by spin-transfer torque, the magnetic field was removed and a pulsed electric current with increased magnitude and fixed pulse width of 100 $\mu$s was applied. The displacement of the DW to the second step could be seen at 4.38 mA as shown in Fig. 4(c).

To move the DW to step 3, an electric current of 4.235 mA was applied (with a small assisted magnetic field of 60 Oe) by keeping the pulse width at 100 $\mu$s [Fig. 4(d)]. The other states shown in Figs. 4(f) and 4(g) could also be obtained under the same magnetic field of 60 Oe and by a small change of the current pulse. When leaving the device for a few hours, we did not see any change of the magnetic state (position of the DW). Importantly, the states discussed above could be obtained again in another experiment under the same conditions demonstrating their reproducibility.

The depinning current $i_{dep}$ is defined as the minimum current magnitude for moving a DW from one state to the other. Both the devices shown in Fig. 5(a) exhibit a dependence of $i_{dep}$ on the applied for a fixed current pulse width of 100 $\mu$s. In the case of the device with $d = 600$ nm and $\lambda = 100$ nm (black dots), $i_{dep}$ is larger but shows a weaker dependence with $H_{per}$. In the case of $d = 200$ nm and $\lambda = 0$ nm, $i_{dep}$ is less and an exponential decay trend is seen (blue dots). This difference in the trend of $i_{dep}$ with the applied magnetic field is worthy of investigation and possibly could be due to the DW configuration in the nano-constriction region.[43,44]

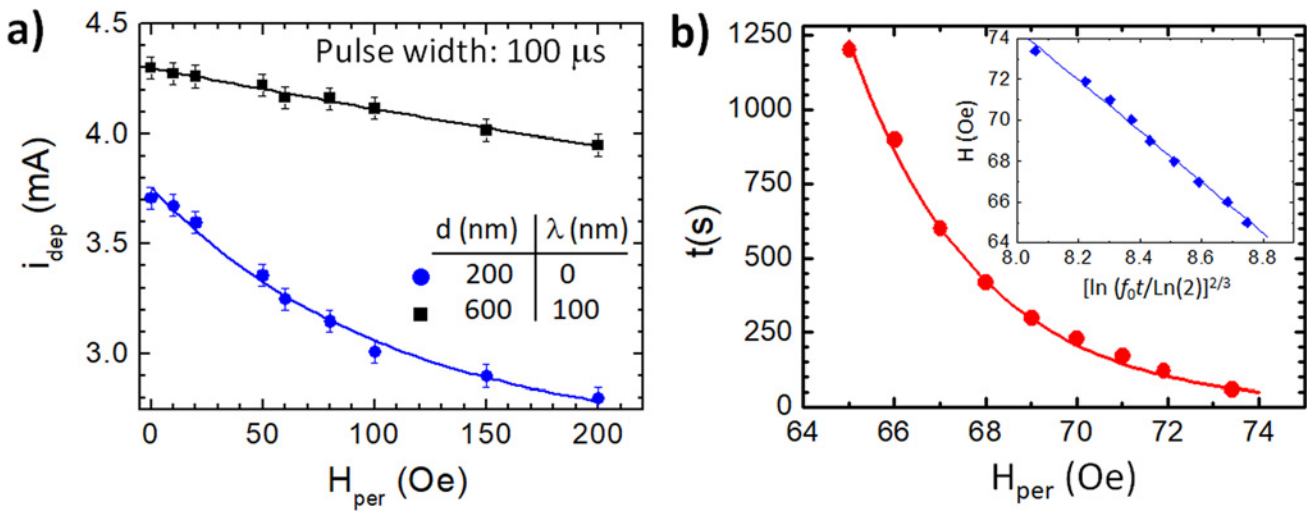

FIG. 5. (a) Depinning current as a function of the applied field for a pulsed current of 100 $\mu$s and (b) time vs applied magnetic field for a device with $d = 600$ nm and $\lambda = 0$ nm. The inset shows the plot of the magnetic field vs time following Eq. (1).

After investigating the possibility of moving a DW between defined positions by an applied electric current, we focused on evaluating its stability. First, the DW was positioned in one step as described above, and by synchronizing the MOKE microscope, the applied current, and the magnetic field, the time before the depinning occurs was measured. Figure 5(b) is a plot of the time, $t$, vs the applied magnetic field for the device with $d = 600$ nm and $\lambda = 0$ nm. It can be noticed that $t$ follows an exponential decay function similar to Sharrock's law,[41] which is applied here to a DW and not the magnetic domain itself. In fact, once the magnetic domain expands or vanishes (becomes unstable), the DW itself will either move or disappear. The applied field vs time can be expressed as[44,45]

$$H = H_0 - B\left[\ln\left(\frac{f_0 t}{\ln(2)}\right)\right]^{2/3}, \tag{1}$$

where $f_0$ is the attempt frequency and $H_0$ and $B$ are the fitting parameters that are related to the stability factor $S\ (= \left(\frac{H_0}{B}\right)^{3/2})$.[46] The best fit to the data shown in the inset of Fig. 5(b) leads to $H_0 = 180.82$ Oe, $B = 13.24$, and $S = 50.12$ using $f_0$ of $1 \times 10^8$ Hz.

In summary, we have demonstrated the controllability of the domain wall position in magnetic wires using spin-transfer torque and a staggered wire design. The depinning current depends on the morphology of the wire. The dependence of depinning current on the applied external magnetic field shows two different behaviors in devices with different $d$. The thermal stability and depinning current can be tuned by adjusting the design parameters of the staggered wires, indicating that this concept offers a high control of the design of a domain wall memory device.

The authors would like to thank S. Al Harthi and M. T. Zar Myint from Sultan Qaboos University for their support and assistance in the magnetometry measurements. S.N.P. acknowledges the partial support from the grant of National Research Foundation, Singapore (NRF2015-IIP003-001).